\documentclass[a4paper,12pt]{article}
\usepackage[utf8x]{inputenc}
\usepackage{amsfonts}
\usepackage{amsmath}
\usepackage{amssymb}
\usepackage{amsthm}
\usepackage{booktabs}
\usepackage{float}
\usepackage{multirow}
\usepackage{graphicx} 
\usepackage{graphics}
\usepackage[round]{natbib}
\usepackage{hyperref}
\usepackage{longtable}
\usepackage[]{threeparttable}
\usepackage{rotating}
\usepackage{xcolor}
\usepackage{comment}

\def\tr{{\rm T}}

\DeclareMathOperator{\logit}{logit}

\def\tr{{\rm tr}}

\newcounter{example}

\title{Median bias reduction in \\  cumulative link models}
\author{V. GIOIA$\,^1$, E. C. KENNE PAGUI$\,^2$ and A. SALVAN$\,^2$ \\
$\,^1\,$\small University of Udine, Department of Economics and Statistics \\
$\,^2\,$\small University of Padova, Department of Statistical Sciences\\
\small gioia.vincenzo@spes.uniud.it, kenne@stat.unipd.it, salvan@stat.unipd.it
}
\date{}
\begin{document}

\maketitle

\begin{abstract}
\noindent
This paper presents a novel estimation approach for cumulative link models, based on median bias reduction as developed in \citet{kenne2017}. The median bias reduced estimator is obtained as solution of an estimating equation based on an adjustment of the score. It allows to obtain higher-order median centering of maximum likelihood estimates without requiring their finiteness. Moreover, the estimator is equivariant under componentwise monotone reparameterizations and the method is effective in preventing boundary estimates. We evaluate the properties of the median bias reduced estimator through simulation studies and compare it with the two main competitors, the maximum likelihood and the mean bias reduced \citep{firth1993} estimators. Finally, we show an application where the proposed estimator is able to solve the boundary estimates problem.

\noindent

\end{abstract}

\noindent
\emph{Some key words:} Adjusted score; Boundary estimate;  Likelihood; Median unbiased;  Ordinal data; Ordinal probability effect measure.

\section{Introduction}
Cumulative link models were proposed by \citet{mccullagh1980}, see also \citet{agresti2010}, and are the most popular tool to handle ordinal outcomes, which are pervasive in many disciplines. One of the reasons for their popularity relies on the use of a single regression coefficient for all response levels, making the effect simple to summarize. For these models, maximum likelihood (ML) is the most common estimation method. Despite  this fact,  it presents some problems and several proposals have been developed  to solve them. One of the problems concerns the asymptotic approximation for the distribution  of the ML estimator, which can be highly inaccurate with  moderate sample information or sparse data.
 Another problem with ML estimation lies in boundary estimates, which can arise with positive probability in models for ordinal data and can cause several difficulties in the fitting process and inferential procedures.

The literature is rich in methods related to bias reduction of the ML estimator. Such methods can be distinguished \citep{kosmidis2014a} into explicit methods, that focus on correcting the estimate, and implicit methods, based on correction of the estimating function. The main disadvantage of the former lies in the need for finiteness of  ML estimates which is overcome by the latter, one of the reasons for their  spread in applied statistics.

The estimation approaches based on an adjustment of the score allow, by introducing an asymptotically negligible bias in the score function, to obtain the mean bias reduced (mean BR) estimator, proposed by  \citet{firth1993} and  developed in  \citet{kosmidis2009, kosmidis2010}, and the  median bias reduced (median BR) estimator, proposed by  \citet{kenne2017}. A unified presentation for generalized linear models is given by  \citet{kosmidis2020} and for general models in  \citet{kenne2019}. Such approaches do not require the finiteness of the ML estimates. In addition, they are effective in preventing boundary estimates. The main difference between the two methods lies in the use of the mean and the median, respectively, as a centering index for the estimator. 
Mean BR achieves a first-order bias correction. The lack of equivariance under nonlinear reparameterizations is a disadvantage of this approach which is, however,  overcome by practical advantages in applications.
Median BR, developed in \citet{kenne2017} and in a subsequent paper \citep{kenne2019}, aims at median centering of the estimator, that is componentwise third-order median unbiased   in the continuous case  and  equivariant under  componentwise monotone reparameterizations.

Mean BR for cumulative link models is developed in \citet{kosmidis2014b}, where finiteness and optimal frequentist properties are illustrated. Here we obtain the quantities needed to compute the median BR in cumulative link models. We use  the simplified algebric form of the adjustment term developed in \citet{kenne2019}. We show, through extensive simulation studies, that the proposed method succeeds in achieving componentwise median centering, outperforms ML and is competitive with mean BR. Considering an ordinal probability effect measure,  proposed by \citet{agrestikateri2017}, we also analyze the behaviour under componentwise monotone reparameterizations, showing the good performance achieved by the median BR estimator. Finally, we present an application where the median BR approach, like mean BR, is seen to be able to prevent boundary estimates. 

\section{Cumulative link models}
Let $Y_i$ be the ordinal outcome, with $c$ categories, for subject $i$, $i=1,\ldots,n$.
Let $p_{ij}=\text{Pr}(Y_i = j)$ be the probability to observe category $j$, $j=1, \ldots, c-1$, for subject $i$,   and $\text{Pr}(Y_i \leq j)=\sum_{k=1}^{j}{p_{ik}}$  the cumulative probability. With $\boldsymbol{x}_i$, $i=1,\ldots,n$, a $p$-dimensional row vector of covariates, the cumulative link model  \citep{mccullagh1980} links the cumulative probabilities to a linear predictor, $\eta_{ij}=\alpha_j+\boldsymbol{x}_i\beta$, $j=1, \ldots, c-1$, via the relationship
\begin{equation}\label{clm}
g\{\text{Pr}(Y_i\leq j|\boldsymbol{x}_{i})\}= \eta_{ij},
\end{equation}
where $g(\cdot)$ is a given link function and  $\beta^\top=(\beta_1, \ldots, \beta_p)$ is the regression parameter vector. This class of models assumes that the effects of $\boldsymbol{x}_{i}$, expressed through $\beta$, are the same for each $j=1, \ldots, c-1$. The intercept parameters $\alpha_j$, $j=1,\ldots, c-1$, satisfy $-\infty=\alpha_0\leq\alpha_1 \leq \ldots \leq \alpha_{c-1}\leq \alpha_c=+\infty$, since $\text{Pr}(Y_i \leq j)$ is increasing in $j$ for each fixed $\boldsymbol{x}_{i}$. Model (\ref{clm}) has an interpretation in terms of an underlying latent variable  \citep[see e.g.][Section 3.3.2]{agresti2010}, that is the ordinal outcome $Y_i$ can be seen as the discretization of a latent continuous random variable $Y^*_i$, satisfying a regression model $Y^*_i=-\boldsymbol{x}_{i} \beta +\varepsilon_i$, $i=1, \ldots, n$. The random variables $\varepsilon_i$ are independent and identically distributed with $E(\varepsilon_i)=0$ and cumulative distribution function $G(\cdot)$. By assigning threshold values $\alpha_j$, $j=1,\ldots,c$, such that we observe $Y_i=j$ if $\alpha_{j-1} \leq Y^*_i < \alpha_j$,  with $-\infty=\alpha_0\leq\alpha_1 \leq \ldots \leq \alpha_{c-1}\leq \alpha_c=+\infty$, the equivalent formulation of model (\ref{clm}) is obtained
\begin{equation*}
\text{Pr}(Y_i\leq j|\boldsymbol{x}_{i})=\text{Pr}(Y^*_i\leq \alpha_j|\boldsymbol{x}_{i})=\text{Pr}(\varepsilon_i<\alpha_j+\boldsymbol{x}_{i}\beta)=G(\eta_{ij}), 
\end{equation*}
with $j=1, \ldots, c-1$. Common choices for $G(\cdot)$ are the logistic, standard normal or extreme value distribution. The cumulative logit model, also known as proportional odds model \citep[Section 2]{mccullagh1980}, is obtained assuming $G(\eta_{ij})=\exp(\eta_{ij})/\{1+\exp(\eta_{ij})\}$, the cumulative probit model is recovered with $G(\eta_{ij})=\Phi(\eta_{ij})$, and the  cumulative complementary log-log link model, also known as proportional hazards model \citep[Section 3]{mccullagh1980}, setting $G(\eta_{ij})=1-\exp\{-\exp(\eta_{ij})\}$.

The popularity of model (\ref{clm}) is linked to its parsimony since it uses a single parameter for each predictor, in addition to the latent variable interpretation. The cumulative link model can be inadequate because of misspecification of the linear predictor or due to departure from the assumption that the covariate effect is the same for each $j$, $j=1,\ldots, c-1$. Several models have been proposed that relax the latter assumption \citep[for a detailed description see][]{fullerton2016}. Instances are the partial cumulative link model, which first appeared in the literature as partial proportional odds model \citep{peterson1990}, or the nonparallel cumulative link model.  Both include the cumulative link model as a special case. However,  despite their flexibility, they may present some difficulties either from a computational or from the interpretation point of view, especially  with data sets with several predictors.

\subsection{Maximum likelihood, bias reduction and boundary estimates}
As the sample size increases, the probability of unique ML estimates tends to one \citep[Section 6.3]{mccullagh1980}. However, the ML estimator has a positive probability of being on the boundary of the parameter space. In cumulative link models (\ref{clm}), boundary estimates are estimates of the regression parameters with infinite components, and/or  consecutive  intercept estimates having the same value. \citet{pratt1981} showed that zero counts for a middle category $j$, $j=2,\ldots,c-1$, produce consecutive equal intercept estimates, that is $\hat \alpha_{j-1}=\hat \alpha_j$, and if the first or the last category have zero observed counts, then the estimates for $\alpha_1$ or  $\alpha_{c-1}$ are infinite. \citet[][Section 3.4.5]{agresti2010} describes some settings where infinite ML estimates  occur for the regression parameters. 

\citet{kosmidis2014b} demonstrates that meanBR is a general effective strategy to prevent boundary estimates. The same advantage will be seen to hold for median BR in Sections 4 and 5. With particular regard to  boundary estimates of the intercept parameters, \citet[Section 8.3, Remark 1]{kosmidis2014b} showed that the ML estimate of the regression parameters is invariant with respect to grouping of unobserved categories with the adjacent ones. So, likelihood inference on the regression parameters is possible if one  or more categories are unobserved. The same appears to hold for mean BR and will be seen to hold in all examples considered for median BR. The only difference with respect to ML estimates is that  if the first or the last category has zero counts, then the mean and median BR estimates are tipically finite. 

\subsection{An ordinal probability effect measure}
A useful monotone transformation of regression parameters related to binary covariates was proposed by \citet{agrestikateri2017} to overcome the difficulty for practitioners to interpret nonlinear measures, such as probits and odds ratios. This reparameterization allows an interpretation in terms of ``ordinal superiority'', that is the probability that an observation from one group falls above an independent observation from the other group, adjusting for other covariates. 
For a vector of covariates $\boldsymbol{x}=(x_1,\ldots,x_p)$, let $x_r$ a binary variable which is a group indicator for an observation. Let $Y_{i1}$, $Y_{i2}$ be the independent outcomes from the groups $x_{ir}=0$ and $x_{ir}=1$, respectively.
For ordinal responses, the ordinal superiority measure, $\gamma \in [0,1]$, is defined as  $$\gamma=\text{Pr}(Y_{i1}>Y_{i2}|\boldsymbol{x}_i \setminus \{x_{ir}\})+\frac{1}{2}\text{Pr}(Y_{i1}=Y_{i2}|\boldsymbol{x}_i \setminus \{x_{ir}\}).$$ 
Based on model (\ref{clm}), \citet{agrestikateri2017} show that the exact or approximate expressions of $\gamma$ for the parameter related to the binary covariate, $\beta_r$, are $\gamma(\beta_r)\approx \exp(-\beta_r/\sqrt 2)/\{1+\exp(-\beta_r/\sqrt 2)\}$, considering the logit link function, $\gamma(\beta_r)=\Phi(-\beta_r/\sqrt 2)$ for the probit link, and  $\gamma(\beta_r)= \exp(-\beta_r)/\{1+\exp(-\beta_r)\}$ for the complementary log-log link.

\section{Median bias reduction}
For a regular parametric model with $p$-dimensional  parameter $\theta=(\theta_1,\ldots,\theta_p)$, let $\ell(\theta)$ be the  log-likelihood  based on a sample of size $n$ and  $U_r=U_r(\theta)=\partial \ell(\theta) / \partial \theta_r$, $r=1,\ldots,p$, the $r$-th component of the score $U(\theta)$. 
Moreover, let $j(\theta)=-\partial^2 \ell(\theta)/\partial \theta\partial \theta^\top$ be the observed information matrix and  $i(\theta)=E_\theta\{j(\theta)\}$ the expected information matrix,  which we assume to be of order $O(n)$. We denote with $[i(\theta)^{-1}]_r$ the $r$-th column of $i(\theta)^{-1}$ and with $i^{rr}(\theta)$ the  $(r,r)$ element of $i(\theta)^{-1}$.

The median BR estimator, $\tilde\theta$, is obtained as solution of the estimating equation $\tilde U(\theta)=0$, where
\begin{equation}\label{adjscore}
\tilde U(\theta)=U(\theta)+\tilde A(\theta),
\end{equation}
with 
\begin{equation*}
\tilde A(\theta)=A^*(\theta)-i(\theta)F(\theta).
\end{equation*}
The vector $A^*(\theta)$ has components $$A^*_r= \frac{1}{2}\tr\{i(\theta)^{-1}(P_r+Q_r)\},$$ with $P_r=E_\theta\{U(\theta)U(\theta)^\top U_r\}$ and $Q_r=-E_\theta\{j(\theta) U_r\}$, $r=1,\ldots, p$. The vector $F(\theta)$ has  components $F_r=[i(\theta)^{-1}]_r^\top \tilde F_r$, where $\tilde F_r$ has elements $$\tilde F_{r,t}= \tr[h_r\{(1/3)P_t+(1/2)Q_t\}],\hspace{0.9cm} r,t=1,\ldots,p,$$ with the matrix $h_r$ given by $$h_r=\frac{[i(\theta)^{-1}]_r[i(\theta)^{-1}]^\top _r}{i^{rr}(\theta)}, \hspace{0.9cm} r=1,\ldots,p.$$  
We refer to \citet{kenne2019} for further details about the computation of $\tilde A(\theta)$ and for the relation with the mean BR estimator \citep{firth1993}, $\hat \theta^*$. The latter is seen to be based on an adjusted score of the form (\ref{adjscore})  with $\tilde A(\theta)=A^*(\theta)$.

\citet{kenne2017} show that in the continuous case, each component of $\tilde \theta$, $\tilde\theta_r$, $r=1,\ldots,p$, is median unbiased with an error of order $O(n^{-3/2})$,  i.e. $\text{Pr}_{\theta}(\tilde\theta_r\leq\theta_r)=\frac{1}{2}+O(n^{-3/2})$, compared to the ML estimator, which is median unbiased with an error of order $O(n^{-1/2})$. Moreover, the asymptotic distribution of  $\tilde\theta$ is  the same as that of the ML estimator, $\hat\theta$, and of the mean BR estimator, $\hat \theta^*$,  that is $\mathcal{N}_p(\theta, i(\theta)^{-1})$.

The equation $\tilde U(\theta)=0$ is usually solved numerically. Moreover, a finite solution is not always guaranteed.
The numerical solutions of $\tilde U(\theta)=0$ can be obtained by a Fisher scoring-type algorithm, whose $(k+1)$-th iteration is 
\begin{equation}\label{fisherscoring}
\theta^{(k+1)}=\theta^{(k)}+i(\theta^{(k)})^{-1}U(\theta^{(k)})+i(\theta^{(k)})^{-1}\tilde A(\theta^{(k)}),
\end{equation}
which differs from the analogue for the ML estimates only by the addition of the term $i(\theta^{(k)})^{-1}\tilde A(\theta^{(k)})$. We adopt, as a stopping criterion for the algorithm, the condition $|\tilde U_r(\theta^{(k)})|<q$, for every $r=1,\ldots,p$, and we set, as default, $q=10^{-10}$.

The algorithm needs a starting value, $\theta^{(0)}$, whose determination is not trivial and can result in nonconvergence of (\ref{fisherscoring}). When available, the ML estimate, $\hat \theta$, or the mean BR estimate, $\hat \theta^*$, are suitable starting values, which are also able to speed up the convergence.  
We set the starting values following a strategy similar to that used in \citet{christensen2019} for cumulative link models (\ref{clm}).
The starting value for the regression coefficients, $\beta$, is set to zero.  
The intercept parameters, $\alpha_j$, $j=1, \ldots, c-1$, are initialized to $\alpha^{(0)}_j=G^{-1}(j/c)$, where $G(\cdot)$ is 
the cumulative distribution function of the error terms, according to the latent variable interpetation discussed in Section 2.

In order to recognize boundary estimates, we adapt the diagnostics in \citet{lesaffre1989}, identifying
infinite estimates  if their absolute value and the corresponding standard error are greater then some thresholds.
 Categories with zero observed counts are grouped, except when it happens at the extreme categories. 

\section{Simulation study}
We  conducted  a simulation study to assess the performance of the median BR estimator, $\tilde \theta$, in cumulative link models (\ref{clm}). We compare it with the ML, $\hat \theta$, and  mean BR, $\hat \theta^*$,   estimators in terms of empirical probability of underestimation (PU\%), estimated relative (mean) bias (RB\%), and empirical coverage of the 95\% Wald-type confidence interval (WALD\%).

We consider sample sizes, $n=50,100,200$, and different  link functions $g(\cdot)$, namely the logit, probit and complementary log-log (cloglog) link functions. We generate the covariate $x_1$ from a standard Normal, $x_2$ and $x_3$ from  Bernoulli distributions with probabilities 0.5 and 0.8 respectively, and $x_4$ from a Poisson with mean 2.5.
Assuming that the response has three categories, we fit the model 
$$g\{\text{Pr}(Y_i\leq j|\boldsymbol{x}_i)\}=\alpha_j +x_{i1}\beta_1+x_{i2}\beta_2+x_{i3}\beta_3+x_{i4}\beta_4, \hspace{0.9cm} j=1,2;\, i=1,\ldots,n,$$
considering 10,000 replications, with covariates fixed at the observed value and true parameter $\theta_0$. Setting $\theta_0=(-1,2,1,-1,1,-1)$ for the logit link function, we use the approximate relations between the coefficients with different link functions leading to $\theta_0=(-0.6,1.2,0.6,-0.6,0.6,-0.6)$ for the probit link function, and  $\theta_0=(-1.1,1,0.7,-0.7,0.7,-0.7)$ for the complementary log-log link function. 

Table \ref{tab1} contains the numerical results for all link functions considered. Boundary estimates occurred using ML with percentage frequencies 2.82\%, 2.75\% and 2.44\%, with $n=50$,  and 0.08\%, 0.1\% and 0.04\%, with $n=100$, for the logit, probit and complementary log-log link functions, respectively. Instead, mean and median BR estimates are always finite. It appears that the new method proves to be remarkably accurate in achieving median centering and  shows a lower estimated relative bias than ML and comparable with that of the mean BR estimator, as well as a good empirical coverage of the the 95\% Wald-type confidence intervals. The differences between the three estimators are appreciable in  lower sample size settings and become much less pronounced as the sample size increases.

\begin{table}[!htbp]
\caption{Estimation of regression parameters  $\beta=(\beta_1, \beta_2, \beta_3, \beta_4)$. Simulation results  for ML, $\hat \beta$, mean BR, $\hat \beta^*$, and median BR, $\tilde \beta$, estimators. For ML, RB\% and WALD\% are conditional upon finiteness of the estimates}\label{tab1}
\centering
\vspace{0.25cm}
\setlength{\tabcolsep}{4pt} 
\renewcommand{\arraystretch}{1.2} 
\footnotesize
\begin{tabular}{ccccccccccc}
\hline
 & &    \multicolumn{3}{c|}{ $n=50$ }& \multicolumn{3}{|c|}{$n=100$} & \multicolumn{3}{|c}{$n=200$} \\ \hline
Link&$\beta$&   \multicolumn{1}{c}{PU\%} &\multicolumn{1}{c}{RB\%}& \multicolumn{1}{c|}{WALD\%} & \multicolumn{1}{c}{PU\%} &\multicolumn{1}{c}{RB\%}& \multicolumn{1}{c|}{WALD\%}&\multicolumn{1}{c}{PU\%} &\multicolumn{1}{c}{RB\%} &  \multicolumn{1}{c}{WALD\%}\\ \hline
\multirow{12}{*}{logit}   & 
$\hat \beta_1$ &40.94&14.50&94.97&43.46&6.30&94.77&45.83&2.80&94.75\\
&$\hat \beta_2$&55.34&14.90&94.76&54.27&6.60&94.93&52.06&2.50&94.88 \\
&$\hat \beta_3$&44.63&13.50&96.48&46.91&9.10&95.32&47.39&4.60&94.97\\ 
&$\hat \beta_4$&62.99&16.50&95.19&59.19&7.00&94.92&56.22&3.20&95.36 \\  \cline{2-11} 
 &
$\hat \beta^*_1$& 54.14&-0.50&95.94&51.99&-0.20&95.34&51.64&-0.30&95.23  \\
&$\hat \beta^*_2$&48.38&0.90&96.35&49.51&0.60&95.77&48.60&-0.30&95.45\\ 
&$\hat \beta^*_3$&53.01&-0.30&96.96&52.64&-0.50&96.06&51.27&0.00&95.52 \\ 
&$\hat \beta^*_4$&45.71&0.40&94.96&47.47&0.00&95.11&47.89&-0.10&95.35\\ \cline{2-11} 
&
$\tilde \beta_1$&50.83&2.90&95.92&50.05&1.20&95.47&50.01&0.40&95.25\\
&$\tilde \beta_2$&50.12&4.20&95.89&50.67&2.10&95.64&49.62&0.40&95.34 \\ 
 &$\tilde \beta_3$&50.12&8.70&97.03&50.60&2.90&95.97&49.99&1.50&95.39\\ 
 &$\tilde \beta_4$&50.22&4.30&95.54&50.34&1.70&95.25&50.07&0.70&95.51\\  \hline
\multirow{12}{*}{probit}  &
$\hat \beta_1$ &40.31& 14.50&94.12&42.82&6.17&94.21&45.23&2.83&94.41\\
 &$\hat \beta_2$&55.40&14.67&94.26&53.65&6.33&94.62&52.44&2.67&94.61 \\
 &$\hat \beta_3$&45.35&12.67&96.35&46.58&8.50&95.02&47.63&4.17&94.82 \\ 
&$\hat \beta_4$&63.26&15.83&94.16&59.23&6.67&94.56&56.74&3.17&95.20\\\cline{2-11} 
&
$\hat \beta^*_1$&53.79&-0.83&95.56&52.18&-0.33&95.15&51.66&-0.17&94.99\\
&$\hat \beta^*_2$&48.67&0.67&96.06&49.30&0.33&95.65&48.69&-0.17&95.06 \\ 
 &$\hat \beta^*_3$&52.93&-1.33&96.79&52.18&-0.67&95.82&51.58&-0.33&95.45 \\ 
 &$\hat \beta^*_4$&44.93&-0.33&94.87&46.40&-0.17&95.18&47.80&0.00&95.17\\  \cline{2-11} 
&
$\tilde \beta_1$&50.81&2.33&95.54&50.08&1.00&95.01&50.23&0.50&94.89\\
&$\tilde \beta_2$&50.46&3.50&95.71&50.23&1.50&95.49&49.37&0.33&94.99 \\ 
 &$\tilde \beta_3$&50.24&6.00&96.89&50.37&2.33&95.63&50.42&1.17&95.23\\ 
&$\tilde \beta_4$&49.67&3.33&95.36&49.35&1.33&95.35&49.90&0.67&95.36\\  \hline
\multirow{12}{*}{cloglog}  &
$\hat \beta_1$&39.59&15.29&94.07&42.58&7.14&94.47&44.69&3.29&94.89\\
&$\hat \beta_2$&55.42&13.86&94.25&53.82&5.86&94.60&52.85&2.86&94.79\\
&$\hat \beta_3$&46.72&15.57& 95.46&46.31&11.43&95.57&47.27&5.86&95.33  \\  
&$\hat \beta_4$&62.53&16.00&94.23&59.16&7.14&94.87&56.04&3.29&95.11\\\cline{2-11} 
&
$\hat \beta^*_1$&55.26&-1.14&95.36&53.07&-0.29&94.89&52.19&-0.29&95.04\\
& $\hat \beta^*_2$&48.95&0.57&96.09&49.17&0.00&95.53&49.46&0.00&95.21\\ 
 & $\hat \beta^*_3$&54.39&-0.86&95.83&52.99&-0.43&95.86&52.02&0.14&95.73\\
&$\hat \beta^*_4$&44.90&0.29&94.73&47.13&0.14&94.94&47.32&0.00&95.37\\\cline{2-11} 
&
$\tilde \beta_1$&51.31&2.57&95.40&50.33&1.43&95.01&50.28&0.71&95.07\\
& $\tilde \beta_2$&50.55&3.43&95.72&50.20&1.29&95.33&50.25&0.71&95.12\\
 & $\tilde \beta_3$&50.77&12.14&96.04&50.16&4.71&95.86&50.10&2.57&95.69\\
& $\tilde \beta_4$&49.95&4.14&95.29&50.73&2.00&95.17&49.52&0.86&95.50\\  \hline
\end{tabular}
\end{table}

 Table \ref{tab2} shows the estimated relative bias under monotone reparameterizations of the parameters related to the binary covariates, considering the ordinal probability effect measure  presented in Section 2.2. In the new parameterization, it appears that the median BR estimator has the best performance in terms of estimated relative bias, if compared with ML and mean BR, which is  not equivariant under this type of reparameterization.

\begin{table}[!htbp] 
\caption{Estimated relative bias (RB\%) for $\gamma(\beta_2)$ and $\gamma(\beta_3)$. For ML, RB\% is conditional upon finiteness of the estimates }\label{tab2}
\centering
\vspace{0.25cm}
\medskip
\setlength{\tabcolsep}{4pt} 
\renewcommand{\arraystretch}{1.2} 
\begin{tabular}{cccccccccc}
link& $n$ & &$\gamma(\hat \beta_2)$&$\gamma(\hat \beta^*_2)$&$\gamma(\tilde \beta_2)$& &$\gamma(\hat \beta_3)$ &$\gamma(\hat \beta^*_3)$&$\gamma(\tilde \beta_3)$ \\
\midrule
\multirow{3}{*}{logit} &$50$&&1.58 &-1.05 &-0.42&& -1.30 &4.15 &1.21\\
&$100$ &&0.79 &-0.49  &-0.18&& -1.70&2.27 &0.88\\ %
&$200$ &&0.24     &-0.39 &-0.22&& -1.00 &1.03  &0.33\\ \hline
\multirow{3}{*}{probit} &$50$&&1.99 &-0.74 &-0.18&&-2.23&-3.43 &0.80\\
&$100$ && 0.93&-0.36  &-0.09&& -2.09&1.73 &0.48\\ %
&$200$ &&0.38      &-0.26&-0.14&& -1.10 &0.80  &0.21\\  \hline
\multirow{3}{*}{cloglog} &$50$&&1.39 &-1.11 &-0.55&& -1.18 &5.18 &1.30\\
&$100$ &&0.63 &-0.61&-0.33&& -2.11&2.59 &0.54\\ %
&$200$ &&0.33 &-0.30&-0.16&& -1.36 &1.12  &0.06\\
\bottomrule[0.09 em]
\end{tabular}
\end{table}

\section{Application}
We consider  the data analysed in \citet{randall1989}, related to a factorial experiment for investigating the factors that affect the bitterness of wine. There are two factors, temperature at the time of crashing the grapes, $x_1$, and contact between juice and skin, $x_2$. Each factor has two levels, ``cold'' and ``warm'' for temperature and ``yes'' and ``no'' for contact. 
For each of the four treatment conditions, two bottles were assessed by a panel of nine judges, giving $n=72$ observations. As in \citet[Section 4.8]{christensen2019}, we consider the outcomes obtained by combining the three central categories and we fit the model 
$$ \logit\{\text{Pr}(Y_i \leq j|\boldsymbol {x}_i)\}=\alpha_j+x_{i1}\beta_1+x_{i2}\beta_2, \hspace{0.9cm} j=1,2;\, i=1,\ldots,72.$$
Table \ref{tab3} shows the coefficient estimates obtained with ML, mean BR and median BR. Both mean and median BR approaches are able to solve the boundary estimates problem.

\begin{table}[!h]
\caption {Coefficient estimates and corresponding standard errors in parenthesis} \label{tab3}
\centering
\vspace{0.25cm}
\setlength{\tabcolsep}{4pt} 
\renewcommand{\arraystretch}{1.2} 
\begin{tabular}{ccccc}
&$\alpha_1$ &$\alpha_2$& $\beta_1$&$\beta_2$\\ \hline
ML&-1.32 (0.53) &$+\infty$ ($+\infty$)&$-\infty$ ($+\infty$) &-1.31 (0.71) \\ 
meanBR &-1.25 (0.51) &5.48 (1.48)&-3.43 (1.42) &-1.19 (0.67)\\ 
medianBR &-1.29 (0.52)& 6.46 (2.32)& -4.48 (2.29) &-1.24 (0.68) \\ 
\bottomrule[0.09 em]
\end{tabular}
\end{table}

Table \ref{tab4} shows the simulation results for the regression parameters  considering 10,000 replications, with covariates fixed at the observed value and true parameter $\theta_0=(-1,4,-2,-1)$. We found $979$ samples out of 10,000 with ML boundary estimates.  Instead, mean and median BR estimates are always finite. The median BR  is again highly accurate in achieving median centering and  shows a lower estimated relative  bias than ML, as well as a good empirical coverage of the 95\% Wald-type confidence intervals.

\begin{table}[!h] 
\caption{Estimation of regression parameters  $\beta=(\beta_1, \beta_2)$. Simulation results  for ML, mean BR and median BR estimators. For ML, RB\% and WALD\% are conditional upon finiteness of the estimates}\label{tab4} 
\centering
\vspace{0.25cm}
\medskip
\setlength{\tabcolsep}{4pt} 
\renewcommand{\arraystretch}{1.25} 
\begin{tabular}{lccccccc}
\toprule[0.09 em]
  &\multicolumn{3}{c}{Parameter $\beta_1$}  & &\multicolumn{3}{c}{Parameter $\beta_2$}\\
\cmidrule{2-4} \cmidrule{6-8} %
 & PU\% & RB\% & WALD\%& & PU\% & RB\% & WALD\% \\
\midrule
ML\tnote{1}  &55.08 & 1.80 &96.92&&53.20  &8.20 &96.50\\
meanBR &43.91  &-0.65 &95.88&& 48.10&0.50 &96.60\\ 
medianBR & 49.71   &8.95  &96.48 && 50.35 &4.90 &96.28\\
\bottomrule[0.09 em]
\end{tabular}
\end{table}

Under the monotone reparameterization  of the coefficients related to the binary covariates, proposed by \citet{agrestikateri2017} and presented in Section 2.2, the estimated percentage relative bias is $-0.81\%$, $1.79\%$ and $0.15\%$ for $\gamma(\beta_1)$,  and $0.69\%$, $0.94\%$ and $0.13\%$ for $\gamma(\beta_2)$, with ML, mean BR and median BR, respectively. For ML, it should be recalled that the estimated relative bias is conditional upon finiteness of the estimates.  It is noteworthy that the median BR estimator has lower estimated relative mean bias that the ML and the mean BR estimators.

\bibliographystyle{chicago}
\bibliography{MBRCLM}
\end{document}